# Comparison of deuterium retention in tungsten exposed to deuterium plasma and gas

Xiaoqiu Ye[a,*], Wei Wang[a], Yifang Wang[a,b], Xiaohong Chen[b], Jiliang Wu[a], Yao Xiao[a],

Xuefeng Wang[a,b], Jun Yan[a,b], Wenzhen Yu[a], Changan Chen[a,*]

[a] Science and Technology on Surface Physics and Chemistry Laboratory, Mianyang 621907, China
[b] School of Sciences and Research Center for Advanced Computation, Xihua University, Chengdu 610039, China

**Abstract**

Deuterium(D) retention behavior in tungsten(W) exposed to deuterium plasma and gas was studied by means of thermal desorption spectroscopy (TDS): deuterium plasma exposure in which W was exposed to D plamsa with 35 eV/$D^+$ at 393 K to the fluence of $3.8 \times 10^{24}$ $D/m^2$; $D_2$ gas charging in which W was exposed to $D_2$ gas of 500 kPa at 773 K for 4 hours. TDS shows that the total D retention in plasma exposure W is ~$10^{22}$ $D/m^2$, one order of magnitude higher than that of gas charging W; however, the $D_2$ desorption peak of gas charging W is 952 K, much higher than 691 K of plasma exposure W. The detrapping energies of deuterium were determined experimentally from the measured peak temperatures at different heating rates and were found to be 2.17 eV for gas charging W and 1.04 eV for plasma exposure W, respectively.

# 1. Introduction

Low tritium (T) retention is one of the main advantages of tungsten(W) as the plasma-facing material (PFM) for future magnetic fusion devices, such as the International Thermonuclear Experimental Reactor (ITER)[1] and the China Fusion

*Corresponding author. Science and Technology on Surface Physics and Chemistry Laboratory, Mianyang 621907, China.
E-mail addresses: xiaoqiugood@sina.com (X.Q. Ye), chenchangan@caep.cn (C.A.Chen).

Engineering Test Reactor (CFETR)[2]. However, during the long-term operation of fusion devices, hydrogen isotpes can easily diffuse deep into W bulk even from the redeposited layers to the W substrate[3]. The tritium retention may increase significantly due to the interaction of thermal diffusion and plasma exposure. Retention and removal of tritium is still one of the most important issues in fusion devices and attract much attention[3-6].

The retention of hydrogen isotopes in W is mainly affected by trap sites, such as dislocations, grain boundaries, vacancies and microvoids in the matrix material[7-10]. It is necessary to evaluate the release behaviors of hydrogen isotopes trapped in different sites. There have been extensive studies on the thermal desorption of deuterium(D) in W by plasma exposure[7, 11-13]. The 'ion-induced' vacancies are usually associated with D agglomeration in molecules and bubbles near the implanted surface. And the voids are often formed from vacancies clustering together to form small microvoids[9]. A single vacancy can trap multi-D atoms, so can voids and bubbles. The detrapping energies of D in vacancies are largely scattered, reported to be 1.0~2.2 eV[7, 9-11, 14-16]. For example, the lower temperature desorption peaks (500-650 K) likely represent single vacancies with a trap energy of 1.0-1.6 eV[11], while the high-temperature desorption peaks (~900 K) are likely associated with the formation of microvoids or vacancy clusters in W with a trap energy of 1.7~2.2 eV[9][11, 14, 16]. Note specially, most of the detrapping energies mentioned above were obtained by fitting numerical calculations based on diffusion-trapping codes to experimental thermal desorption spectra(TDS)[16]. A large uncertainty in determination of characteristics of defects in such approach is given by dependence of TDS spectrum simulation on many input parameters, such as characteristic frequencies for trapping or detrapping, depth distribution of traps and trapped D[11]. The direct experimental values of detrapping energies are still scarce[17]. Zibrov et al. made these efforts[11, 17]. The D detrapping energies for single vacancies and vacancy clusters was determined experimentaly to be 1.56 eV and 2.10 eV, respectively[11, 17].

On the other hand, as expected, hydrogen isotopes will also enter into W by thermal diffusion during the process of continuous plasma bombardment[18]. Compared with plasma exposure W, gas charging W can avoid the influence of surface damage layer, and is more conducive to the understanding of the intrinsic mechanism of deuterium retention in W by thermal diffusion. Moreover, due to the deeper diffusion depth in W, this part of hydrogen isotpes may be hard to remove.

This work is devoted to the study on the D retention behavior of W samples by two treatment processes, respectively: D plasma exposure in which W was exposed to D plamsa with 35 eV/$D^+$ at 393 K to the fluence of $3.8\times10^{24}$ D/$m^2$ (the sample was denoted as plasma exposure W); $D_2$ gas charging in which W was exposed to $D_2$ gas of 500 kPa at 773 K for 4 hours (the sample denoted as charging W). The amount, desorption temperatures and detrapping energies of deuterium in W exposure to D plasma and $D_2$ gas were determined experimentally.

## 2 Experimental

Rolled W (purity of >99.95%) purchased from Advanced Technology & Materials Co., Ltd. was used in this study. The samples with dimensions of $10\times 10\times 1$ $mm^3$ were all mechanically polished to obtain a mirror-like surface, and then ultrasonically cleaned in ethanol and acetone. Thereafter, high-temperature annealing (1273 K/1 h) was performed in a vacuum of better than $10^{-5}$ Pa for outgassing. The sample has a typical microstructure with the average grain size of about 4 μm (details are given in [7]).

The schematic diagram of the gas charging system was shown in Fig. 1. $D_2$ was stored in a ZrCo storage bed, which was heated to drive $D_2$ into the standard vessel. The pipe system was cleaned three times by 10 kPa $D_2$ before each experiment. The samples were exposed to $D_2$ with a pressure of 500 kPa at 773 K for 4 h to obtain D saturated W samples according to the diffusive transport parameters of $D_2$ through W samples[19].

After exposure, all samples were cooled rapidly to room temperature in deuterium atmosphere. Note specially, the effect of deuterium desorption during cooling is not considered here, since all the samples have this same process[20, 21]. Subsequently TDS experiments were followed. The 316 L stainless steel pipe which ions passed through was baked for 24 h to remove residual gas in the materials before TDS experiment. The samples were heated linearly up to 1273 K in the quartz tube, with a vacuum of $10^{-5}$ Pa. In addition, in order to calculate the activation energy of $D_2$ desorption peak, the heating rate was set to be 5 K/min, 10 K/min, 15 K/min and 20 K/min respectively for the D saturated W samples. $D_2$ and HD signals were tracked by quadrupole mass spectrometer (QMS). And the $D_2/H_2$ signal was calibrated using two standard leaks after experiments. HD signal could be theoretically calibrated, which was the average of the sum of $D_2$ and $H_2$[20].

.

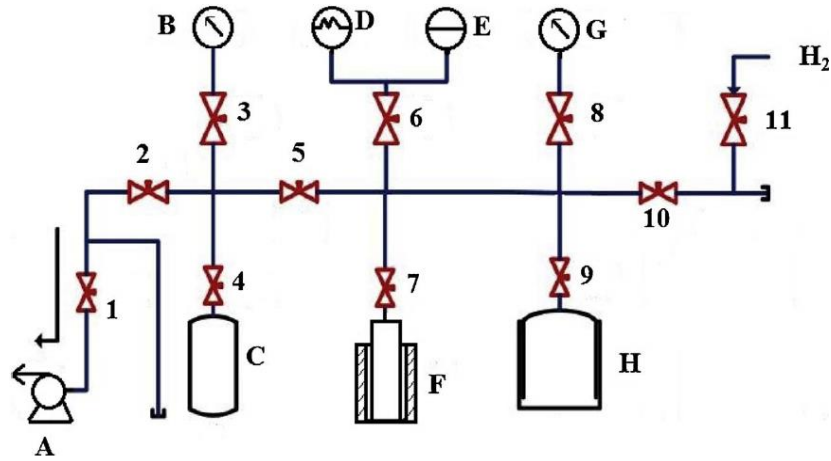

**Fig. 1.** Schematic diagram of the gas charging system. A: scroll pump, B: pressure gauge (∼1 bar), C: standard vessel, D: resistive vacuum gauge, E: film vacuum gauge, F: heating apparatus, G: pressure gauge (∼10 bar), H: Zr-Co alloy storage bed, 1–11: Swagelok valves.

Deuterium plasma exposure was performed for W in the linear experimental plasma system (LEPS) at Lanzhou Institute of Chemical Physics, Chinese Academy of Sciences. During the exposure, ion energy was set at 35 eV/$D^+$, and the exposure temperature was kept at around 393 K. The ion flux which consisted dominantly of $D_3^+$ ions[22] was around $2.1 \times 10^{21}$ D/$m^2$ s. Thus, the fluence was $3.8 \times 10^{24}$ D/$m^2$ for exposure periods of 0.5 h. TDS was also used to get data on D trapping and retention in plasma exposure W.

# 3 Results and discussion

## 3.1 Desorption Characteristics of deuterium

Fig.2 shows the desorption rate of $D_2$ for plasma exposure W and gas charging W. It is clear that the $D_2$ desorption peak of gas charging W is about 952 K, much higher than 691 K of plasma exposure W. Moreover, the temperature range of significant $D_2$ release for gas charging W is extending from 600 to 1173 K; while for the plasma exposure W, this range is 550 to 800 K. Deuterium in gas charging W is more difficult to release than that in plasma exposure W. It suggests that there are different D trap sites in these two kinds of samples.

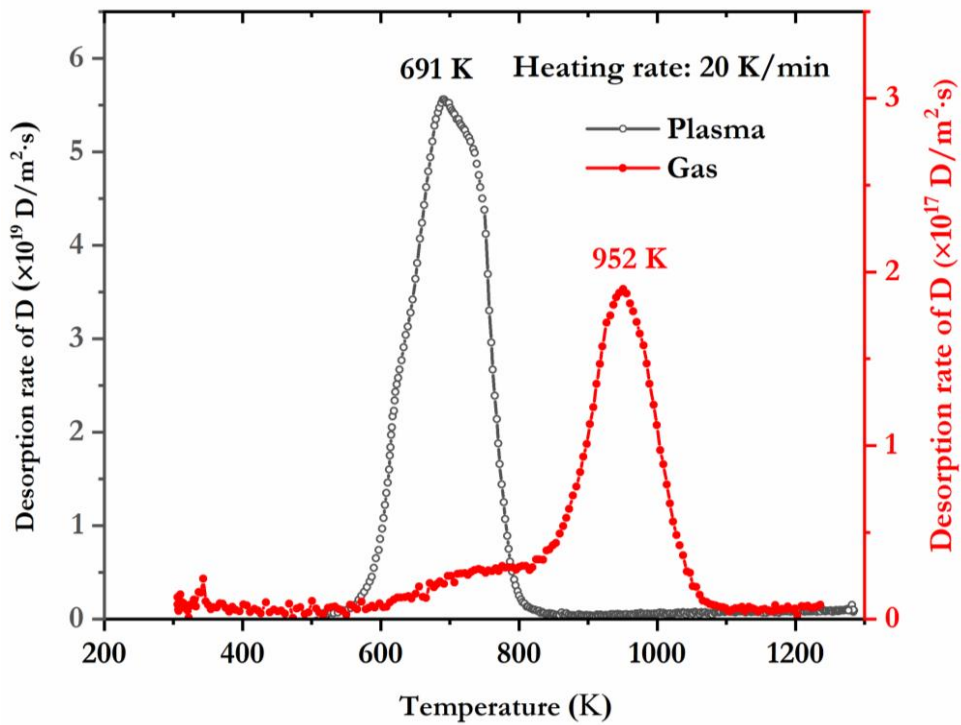

**Fig.2**. Desorption rate of $D_2$ for plasma exposure W and gas-charging W. During the plasma exposure, deuterium ion energy, beam flux, fluence and irradiation temperature is 35 eV/$D^+$, $2.1 \times 10^{21}$ D/($m^2$ s), $3.8 \times 10^{24}$ D/$m^2$ and 393 K, respectively; Deuterium gas pressure, sample temperature and holding time are 500 kPa, 773 K and 4 h, respectively during gas charging.

As the literatures states [11, 23], the peak at around 690 K in plasma exposure W, may correspond to the D release from single vacancies[11], while the peak at around 950 K in gas charging W may correspond to the D release from vacancy clusters [11]or microvoids[23]. In addition, the spectra for 952 K have a shoulder at the temperature of 700 K that is presumably corresponding to the D release from single vacancies. We will discuss this further in the next session.

It is worth mentioning that although the deuterium injection layer is very shallow(about 4 nanometers from SRIM[24] simulation). However, the amount of deuterium retention by plasma exposure is $1.90 \times 10^{22}$ D/m$^2$, nearly one order of magnitude higher than that by gas charging, as shown in Fig.3. The origin of such diferences is not completely clear. It may be due to that the deuterium entering into W by plasma exposure is supersaturated[3, 25]. For example, Gao et al.[25] even found the existence of a 10 nm thick D-supersaturated surface layer (DSSL) with an unexpectedly high D concentration of ~10 at.% after irradiation with ion energy of 215 eV at 300 K[25].

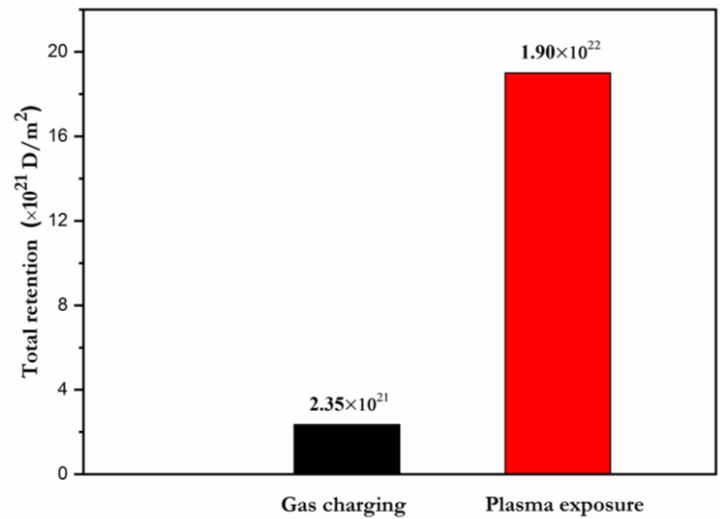

**Fig. 3.** Comparison of total D retention for plasma exposure W and gas charging W

The comparison of deuterium retention for plasma exposure W and gas charging W indicates that the non-equilibrium process of deuterium entering into W by plasma exposure may be the main component of deuterium retention in tungsten, and near-surface vacancies are responsible for this retention[26].

The measured deuterium content in gas charging W is the order of $10^{21}$ D/m$^2$,

much higher than the solubility of interstitial deuterium in W[21, 27]. This difference also indicates that defects (e.g. dislocations, grain boundaries and vacancies) would play a dominant role in the actual amount of deuterium retention in W [28]. H loaded over solubility limits should stay somewhere in the bulk accompanying some lattice distortion and also making bubbles[3], such as H-induced superabundant vacancies[29]. The peak of 952 K in gas charging W is also in accordance with the characteristics of the D release from vacancy clusters [11].

**3.2 Detrapping energies of deuterium**

Fig.4 shows the TDS spectra of $D_2$ in plasma exposure W and gas charging W at different heating rates. It can be seen that the desorption peaks of deuterium in W moves to higher temperatures with the increase of heating rate. For example, when the heating rate is 5 K/min, 10 K/min, 15 K/min and 20 K/min, the desorption peaks of gas-charging W locate at 909, 925, 943 and 952 K , respectively; while the desorption peaks of plasma exposure W locate at 646, 670, 683 and 691 K, respectively. The desorption peak temperatures of deuterium in plasma exposure W were generally lower than that in gas charging W at the same heating rate. This further confirms that deuterium in gas charging W is more difficult to be removed.

The detrapping energy means the energy barrier for D escape from a trap, which is usually defined as the sum of the bingding energy and the activation energy for D diffusion[17]. Deuterium trapped by various types of defects in tungsten requires different detrapping energies. The literature states that desorption peaks near 900 K correspond to trap energies of ~2.1 eV[16]. These traps are typically voids [30]or vacancy clusters[11].  The peaks at around 600 K   correspond to the D release from single vacancies with  trap energies ranging from 1.1 to 1.4 eV [11, 31]. Thus, it is interesting to determine the detrapping energies of D in gas charging W and plasma exposure W.

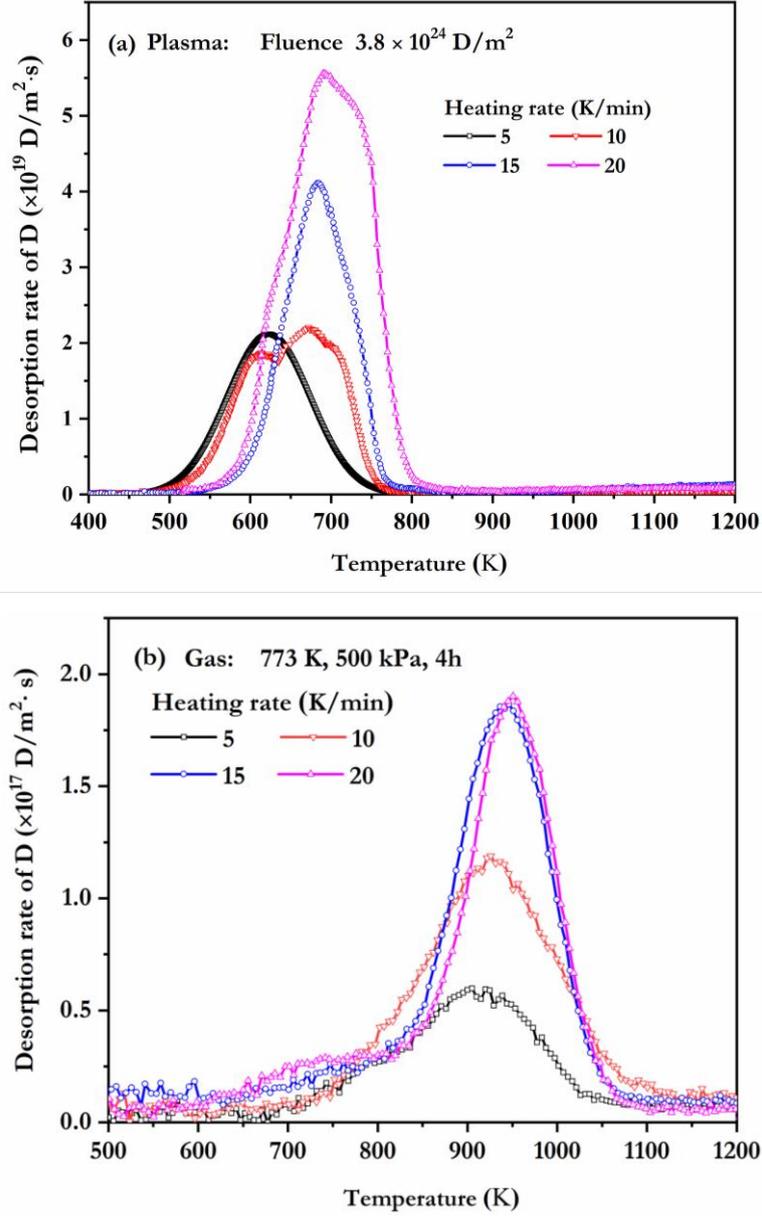

**Fig.4.** Thermal desorption spectra of $D_2$ from plasma exposure W (a) and gas charging W(b) at different heating rates.

The detrapping energies of trapped deuterium can be determined experimentally from the measured peak temperatures at different heating rates[11, 32], as described by the following equation:

$$\ln(\frac{\beta}{T_p^2}) = \ln(A\frac{R}{E_{dt}}) - \frac{E_{dt}}{R}\frac{1}{T_p} \tag{1}$$

where $E_{dt}$ is the detrapping energy for D in the metal, $\beta$- the heating rate, $T_p$ - the temperature corresponding to the TDS peak of interest, A - a constant depending on parameters of the material and trapping sites, and R-the universal gas constant.

Since $\beta$ and $T_p$ are known, the detrapping energy $E_{dt}$ of deuterium evolution from a trapping site can be calculated from the slope of a $\ln(\beta/T_p^2)$ vs $(1/T_p)$ plot. The results are shown in Fig. 5. The fitting results have good linear relationships and satisfy the Arrehenius relationship[32]. The detrapping energies of deuterium in gas charging and plasma exposure samples can be calculated as 2.17 eV and 1.04 eV, respectively.

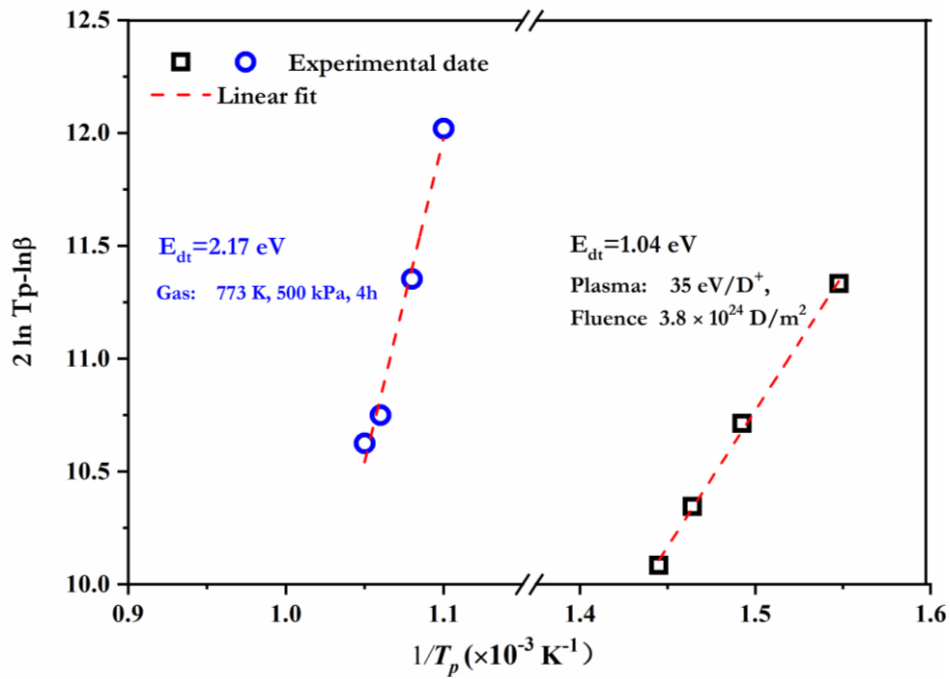

**Fig. 5.** The plot of $[-\ln(\beta/T_p^2)]$ versus $1/T_p$ for the main peaks of plasma exposure W and and gas charging W. The best linear fit and the determined value of Edt is also shown.

The detrapping energy of deuterium in gas charging samples is 2.17 eV, which is close to the value of 2.10 eV that was interpreted by several others as trapping of deuterium in vacancy clusters[11] or voids [9, 16]. The measured deuterium content in gas charging samples is much higher than deuterium solubility in tungsten [21]. It also indicates that vacancy clusters or voids may play a dominant role in the actual amount of deuterium retention in tungsten[33]. Note specially, these high-energy traps in W are usually formed by irradiation with neutron[9], high energy ions (such as 10 keV/D

ions[11]), or by the plasma exposure at high surface temperatures（680-950 K）of W[9], while they are seldom reported to be found in the gas charging W.

The positron anihilation spectroscopy (PAS) measurements confirmed that monovacancy is usually found to be dominant in the undamaged W[7]. Considering that W was exposed to $D_2$ with a pressure of 500 kPa at 773 K for 4 h in this work, this results further comfirmed that H isotopes could enhance vacancy formation [34]. Vacancy clustering or voids may also easily take place due to the temperature higher than 550 K [9, 23, 35]. Thus, the high-temperature desorption peak (950 K) in the gas charging sample is likely associated with the formation of microvoids or vacancy clusters in the W with a trap energy of ~2.17 eV.

The detrapping energy of the sample irradiated by plasma is 1.04 eV, which is in agreement with the trapping energy of deuterium by a single vacancy as reported by Poon[16] (~1.07 eV). It is worth noting that although the amount of deuterium retention by gas charging is much lower than that by plasma exposure, higher trap energies and physically deeper trap location could all increase the temperature of the release peaks of D in the gas charging sample[16].

**4 Conclusions**

We carried out a study on the amount, desorption temperatures and detrapping energies of deuterium in tungsten(W) exposure to D plasma and $D_2$ gas. The results showed that the total D retention in plasma exposure W is ~$10^{22}$ $D/m^2$, one order of magnitude higher than that of gas charging W. Whereas, the $D_2$ desorption peak of gas charging W is around 950 K, higher than that of plasma exposure W. The detrapping energies of deuterium in gas charging W and plasma exposure W were found to be 2.17 eV and 1.04 eV, respectively. Higher trap energy and physically deeper trap location could all increase the temperature of the release peaks of D in the gas charging sample. These results also suggests that deuterium in gas charging W was likely trapped by vacancy clusters while deuterium in plasma exposure W was trapped by a single vacancy.

## Acknowledgements

We are grateful to Zhanlei Wang, Yuwei You, Xuexi Zhang, Peng Wang and Guanghong Lv for their discussions. We acknowledge support by National Magnetic Confinement Fusion Energy Research Project (No. 2015GB109002) from Ministry of Science and Technology of China. The authors also appreciate the support from the National Nature Science Foundation of China (No. 21401173).

## References

[1] R. A. Pitts, S. Carpentier, F. Escourbiac, T. Hirai, V. Komarov, S. Lisgo, A. S. Kukushkin, A. Loarte, M. Merola, A. Sashala Naik, R. Mitteau, M. Sugihara, B. Bazylev, P. C. Stangeby, A full tungsten divertor for ITER: Physics issues and design status, J. Nucl. Mater., 438(2013) S48-S56, https://doi.org/10.1016/j.jnucmat.2013.01.008.

[2] Y. Wan, J. Li, L. Yong, X. Wang, Q. Li, Overview of the present progress and activities on the CFETR, Nucl. Fusion, 57(2017) 102009, https://doi.org/10.1088/1741-4326/aa686a.

[3] T. Tanabe, Review of hydrogen retention in tungsten, Phys. Scripta, T159(2014) 014044, https://doi.org/10.1088/0031-8949/2014/T159/014044.

[4] K. Tokunaga, M. J. Baldwin, R. P. Doerner, N. Noda, Y. Kubota, N. Yoshida, T. Sogabe, T. Kato, B. Schedler, Blister formation and deuterium retention on tungsten exposed to low energy and high flux deuterium plasma, J. Nucl. Mater., 337-339(2005) 887-891, https://doi.org/10.1016/j.jnucmat.2004.10.137.

[5] V. K. Alimov, J. Roth, Hydrogen isotope retention in plasma-facing materials: review of recent experimental results, Phys. Scripta, T128(2007)6-13, http://doi:10.1088/0031-8949/2007/T128/002.

[6] R. Joachim, T. Emmanuelle, L. Thierry, P. Volker, B. Sebastijan, L. Alberto, F. C. Glenn, P. D. Russell, S. Klaus, V. O. Olga, A. C. Rion, Tritium inventory in ITER plasma-facing materials and tritium removal procedures, Plasma Phys. Contr. F., 50(2008) 103001, https://doi.org/10.1088/0741-3335/50/10/103001.

[7] S. Wang, X. Zhu, L. Cheng, W. Guo, M. Liu, C. Xu, Y. Yuan, E. Fu, X.-Z. Cao, G.-H. Lu, Effect of heavy ion pre-irradiation on blistering and deuterium retention in tungsten exposed to high-fluence deuterium plasma, J. Nucl. Mater., 508(2018) 395-402, https://doi.org/10.1016/j.jnucmat. 2018.05.082.

[8] O. El-Atwani, C. N. Taylor, J. Frishkoff, W. Harlow, E. Esquivel, S. A. Maloy, M. L. Taheri, Thermal desorption spectroscopy of high fluence irradiated ultrafine and nanocrystalline tungsten: helium trapping and desorption correlated with morphology, Nucl. Fusion, 58(2018) 016020, https://doi.org/10.1088/1741-4326/aa86cf.

[9] G. M. Wright, M. Mayer, K. Ertl, G. de Saint-Aubin, J. Rapp, Hydrogenic retention in irradiated tungsten exposed to high-flux plasma, Nucl. Fusion, 50(2010) 075006, https://doi.org/10.1088/0029-5515/50/7/075006.

[10] Y. Oya, K. Azuma, A. Togari, Q. Zhou, Y. Hatano, M. Shimada, R. Kolasinski, D. Buchenauer In, Interaction of hydrogen isotopes with radiation damaged tungsten, Recent Advances in

Technology Research and Education, Springer International Publishing, 2018, 41-49, https://doi.org/10.1007/978-3-319-67459-9_6.

[11] S. Ryabtsev, Y. Gasparyan, M. Zibrov, A. Shubina, A. Pisarev, Deuterium thermal desorption from vacancy clusters in tungsten, Nucl. Instrum. Meth. B., 382(2016) 101-104, https://doi.org/10.1016/j.nimb.2016.04.038.

[12] G. N. Luo, K. Umstadter, W. M. Shu, W. Wampler, G. H. Lu, Behavior of tungsten with exposure to deuterium plasmas, Nucl. Instrum. Meth. B., 267(2009) 3041-3045, https://doi.org/10.1016/j.nimb. 2009.06.049.

[13] M. Zibrov, M. Balden, T. W. Morgan, M. Mayer, Deuterium trapping and surface modification of polycrystalline tungsten exposed to a high-flux plasma at high fluences, Nucl. Fusion, 57(2017) 046004, https://doi.org/10.1088/1741-4326/aa5898.

[14] Y. M. Gasparyan, O. V. Ogorodnikova, V. S. Efimov, A. Mednikov, E. D. Marenkov, A. A. Pisarev, S. Markelj, I. Čadež, Thermal desorption from self-damaged tungsten exposed to deuterium atoms, J. Nucl. Mater., 463(2015) 1013-1016, https://doi.org/10.1016/j.jnucmat.2014.11.022.

[15] J. Guterl, R. D. Smirnov, S. I. Krasheninnikov, M. Zibrov, A. A. Pisarev, Theoretical analysis of deuterium retention in tungsten plasma-facing components induced by various traps via thermal desorption spectroscopy, Nucl. Fusion, 55(2015) 093017, https://doi.org/10.1088/0029-5515/55/9/093017.

[16] M. Poon, A. A. Haasz, J. W. Davis, Modelling deuterium release during thermal desorption of D+-irradiated tungsten, J. Nucl. Mater., 374(2008) 390-402, https://doi.org/10.1016/j.jnucmat. 2007.09.028.

[17] M. Zibrov, S. Ryabtsev, Y. Gasparyan, A. Pisarev, Experimental determination of the deuterium binding energy with vacancies in tungsten, J. Nucl. Mater., 477(2016) 292-297, https://doi.org/10.1016/j.jnucmat.2016.04.052.

[18] T. Otsuka, T. Tanabe, K. Tokunaga, Retention and release mechanisms of tritium loaded in plasma-sprayed tungsten coatings by plasma exposure, J. Nucl. Mater., 438(2013) S1048-S1051, https://doi.org/10.1016/j.jnucmat.2013.01.229.

[19] F. Liu, H. Zhou, Y. Xu, X.-C. Li, M. Zhao, T. Zhang, Z. Xie, Q.-Z. Yan, X. Zhang, F. Ding, S. Liu, G.-N. Luo, Gas-driven permeation of deuterium through tungsten with different microstructures, Fusion Eng. Des., 113(2016) 216-220, https://doi.org/10.1016/j.fusengdes.2016.09.003.

[20] Z. Wang, C. A. Chen, Y. Song, X. Xiang, W. Wang, L. Liu, B. Wang, Deuterium retention removal in China reduced activation ferritic-martensitic steels through thermal desorption and hydrogen isotope exchange, Fusion Eng. Des., 126(2018) 139-146, https://doi.org/10.1016/ j.fusengdes .2017.11.024.

[21] R. Frauenfelder, Solution and diffusion of hydrogen in tungsten, J. Vac. Sci. Technnol., 6(1969) 388-397, https://doi.org/10.1116/1.1492699.

[22] O. V. Ogorodnikova, T. Schwarzselinger, K. Sugiyama, T. Dürbeck, W. Jacob, Deuterium retention in different tungsten grades, Phys. Scripta, T138(2009) 014053, https://doi.org/10.1088/0031-8949/2009/T138/014053.

[23] H. Eleveld, A. van Veen, Void growth and thermal desorption of deuterium from voids in tungsten, J. Nucl. Mater., 212-215(1994) 1421-1425, https://doi.org/10.1016/0022-3115(94)91062-6.

[24] J. F. Ziegler, M. D. Ziegler, J. P. Biersack, SRIM - The stopping and range of ions in matter (2010), Nucl. Instrum. Meth. B., 268(2010) 1818-1823, https://doi.org/10.1016/j.nimb.2010.02.091.

[25] L. Gao, W. Jacob, U. von Toussaint, A. Manhard, M. Balden, K. Schmid, T. Schwarz-Selinger, Deuterium supersaturation in low-energy plasma-loaded tungsten surfaces, Nucl. Fusion, 57(2016) 016026, https://doi.org/10.1088/0029-5515/57/1/016026.

[26] D. H. Karl, Helium, hydrogen, and fuzz in plasma-facing materials, Mater. Res. Express, 4(2017) 104002, https://doi.org/10.1088/2053-1591/aa8c22.

[27] G. A. Esteban, A. Perujo, L. A. Sedano, K. Douglas, Hydrogen isotope diffusive transport parameters in pure polycrystalline tungsten, J. Nucl. Mater., 295(2001) 49-56, https://doi.org/10.1016/S0022-3115(01)00486-X.

[28] X.-S. Kong, S. Wang, X. Wu, Y.-W. You, C. S. Liu, Q. F. Fang, J.-L. Chen, G. N. Luo, First-principles calculations of hydrogen solution and diffusion in tungsten: Temperature and defect-trapping effects, Acta Mater., 84(2015) 426-435, https://doi.org/10.1016/j.actamat.2014.10.039.

[29] K. Ohsawa, F. Nakamori, Y. Hatano, M. Yamaguchi, Thermodynamics of hydrogen-induced superabundant vacancy in tungsten, J. Nucl. Mater., 458(2015) 187-197, https://doi.org/10.1016/j.jnucmat.2014.12.029.

[30] Y. M. Gasparyan, A. V. Golubeva, M. Mayer, A. A. Pisarev, J. Roth, Ion-driven deuterium permeation through tungsten at high temperatures, J. Nucl. Mater., 390-391(2009) 606-609, https://doi.org/10.1016/j.jnucmat.2009.01.172.

[31] M. H. J. t. Hoen, M. Mayer, A. W. Kleyn, H. Schut, P. A. Z. v. Emmichoven, Reduced deuterium retention in self-damaged tungsten exposed to high-flux plasmas at high surface temperatures, Nucl. Fusion, 53(2013) 043003, https://doi.org/10.1088/0029-5515/53/4/043003.

[32] W. Y. Choo, J. Y. Lee, Thermal analysis of trapped hydrogen in pure iron, Metall. Mater. Trans. A, 13(1982) 135-140, https://doi.org/10.1007/BF02642424..

[33] Z. Zhe, Y. Li, C. Zhang, G. Pan, P. Tang, Z. Zhi, Effect of grain size on the behavior of hydrogen/helium retention in tungsten: A cluster dynamics modeling, Nucl. Fusion, 57(2017) 086020.

[34] S. C. Middleburgh, R. E. Voskoboinikov, M. C. Guenette, D. P. Riley, Hydrogen induced vacancy formation in tungsten, J. Nucl. Mater., 448(2014) 270-275, https://doi.org/10.1016/j.jnucmat.2014.02.014.

[35] S. A. Ryabtsev, Y. M. Gasparyan, M. S. Zibrov, A. A. Pisarev, On the annealing of radiation-induced point defects in tungsten, J. Surf. Investig-X-RA, 10(2016) 658-662, https://doi.org/10.1134/S1027451016030332.